\documentclass[conference]{IEEEtran}
\IEEEoverridecommandlockouts
\usepackage{cite}
\usepackage{amsmath,amssymb,amsfonts}
\usepackage{algorithmic}
\usepackage{graphicx}
\usepackage{textcomp}
\usepackage{xcolor}
\usepackage{academicons}
\definecolor{orcidlogocol}{HTML}{A6CE39}
\usepackage{orcidlink}
\usepackage{subcaption}

\usepackage[utf8]{inputenc} 
\usepackage[T1]{fontenc}    
\usepackage{hyperref}       
\usepackage{url}            
\usepackage{booktabs}       
\usepackage{amsfonts}       
\usepackage{nicefrac}       
\usepackage{microtype}      
\usepackage{xcolor}         
\usepackage{amsmath}
\usepackage{amssymb}
\usepackage{graphicx}
\usepackage[table]{xcolor}
\usepackage{colortbl}
\usepackage{hyperref}
\usepackage{float}

\usepackage{caption}
\usepackage{cuted}

\usepackage{caption}

\makeatletter
\def\section{\@startsection {section}{1}{\z@}
    {1.5ex plus 1ex minus .2ex}
    {1ex plus .2ex}
    {\normalfont\bfseries\normalsize\centering}}
\def\subsection{\@startsection{subsection}{2}{\z@}
    {1ex plus 1ex minus .2ex}
    {0.8ex plus .2ex}
    {\normalfont\bfseries\normalsize}}
\def\subsubsection{\@startsection{subsubsection}{3}{\z@}
    {1ex plus 1ex minus .2ex}
    {0.8ex plus .2ex}
    {\normalfont\bfseries\small}}
\makeatother

\def\BibTeX{{\rm B\kern-.05em{\sc i\kern-.025em b}\kern-.08em
    T\kern-.1667em\lower.7ex\hbox{E}\kern-.125emX}}
\begin{document}

\title{Diffusion Models for Joint Audio-Video Generation\\}

\author{\IEEEauthorblockN{Alejandro Paredes La Torre \orcidlink{0009-0002-3303-3442}}
\IEEEauthorblockA{\textit{Duke University}\\
alejandro.paredeslatorre@duke.edu}
}

\maketitle

\begin{abstract}
Multimodal generative models have shown remarkable progress in single-modality video and audio synthesis, yet truly joint audio–video generation remains an open challenge. In this paper, I explore four key contributions to advance this field. First, I release two high-quality, paired audio–video datasets. The datasets consisting on 13 hours of video-game clips and 64 hours of concert performances, each segmented into consistent 34-second samples to facilitate reproducible research. Second, I train the MM-Diffusion architecture from scratch on our datasets, demonstrating its ability to produce semantically coherent audio–video pairs and quantitatively evaluating alignment on rapid actions and musical cues. Third, I investigate joint latent diffusion by leveraging pretrained video and audio encoder–decoders, uncovering challenges and  inconsistencies in the multimodal decoding stage. Finally, I propose a sequential two-step text-to-audio-video generation pipeline: first generating video, then conditioning on both the video output and the original prompt to synthesize temporally synchronized audio. My experiments show that this modular approach yields high-fidelity generations of audio video generation. The code can be found in https://github.com/AlejandroParedesLT/audioVideo-GenAI
\end{abstract}


\begin{figure*}[t]
    \centering
\includegraphics[height=3cm,width=\textwidth]{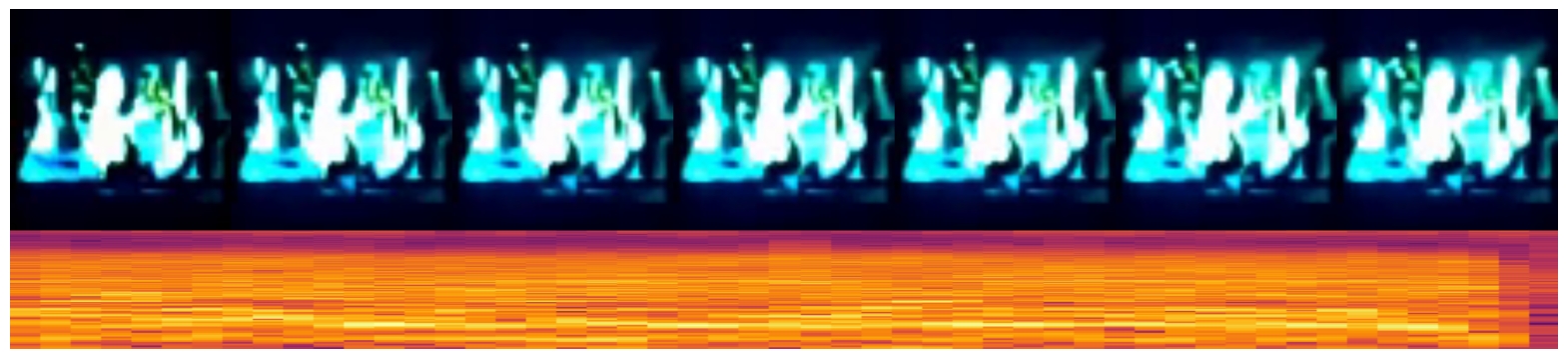}
    \caption{MM-Diffusion unconditional generation. Model trained from scratch on the concert dataset (20k steps). The model is able to capture the semantics of the dataset (lights and human figures). Further training can yield better results.}
    \label{fig:fullwidth_image}
\end{figure*}

\vspace{-1em}
\begin{center}
    \includegraphics[width=\linewidth]{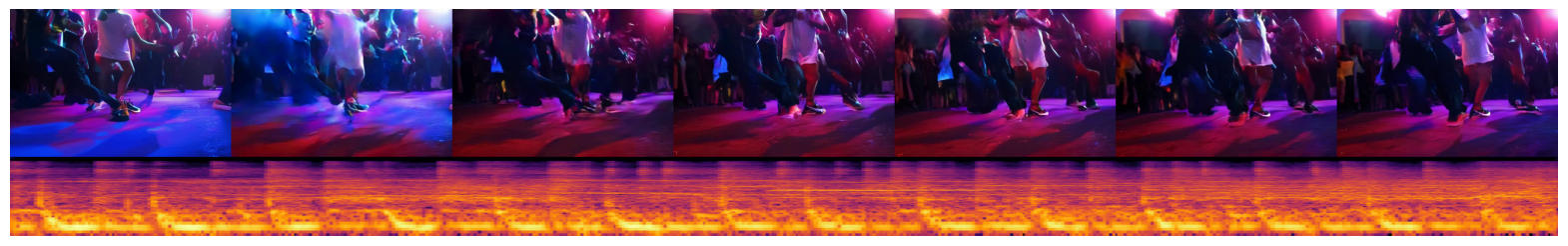}\\[0.5em]
    \captionof{figure}{Two-step sequential generation using prompt "A lively street dance battle under neon lights, with dancers showing off impressive moves to an energetic hip-hop beat. The crowd cheers ...". A clear alignment between the motion of dancing and rhythmic patterns reflected on the spectrogram can be observed.}
    \label{fig:titlepageimage}
\end{center}
\vspace{1em}

\section{Introduction}
Multimodal generative models have recently emerged as a vibrant research area, driven by advances in diffusion techniques \cite{yang2023dawnlmmspreliminaryexplorations,nvidia2025cosmosworldfoundationmodel,liu2024sorareviewbackgroundtechnology}. Research on video generation models has seen general improvements on massive long-context generation, with increasing work being done on foundational models that can effectively generate videos from a very wide range of topics \cite{nvidia2025cosmosworldfoundationmodel, blattmann2023alignlatentshighresolutionvideo}. Multimodal generative models are increasingly gaining attention with the increasing attention to the research of large language models and diffusion models. Building on important contributions for video generation \cite{blattmann2023alignlatentshighresolutionvideo}, powerful models such as Stable Video diffusion \cite{blattmann2023stablevideodiffusionscaling} and Sora \cite{liu2024sorareviewbackgroundtechnology} can create guided text-to-video, based on diffusion techniques extensively used in image generation \cite{ramesh2022hierarchicaltextconditionalimagegeneration}, while models  such as AudioLDM \cite{liu2023audioldmtexttoaudiogenerationlatent}, MusicGen \cite{copet2024simplecontrollablemusicgeneration} and AudioGen \cite{kreuk2023audiogentextuallyguidedaudio} generate audio and music effects using very efficient techniques. 

These models are limited to single-modality generation, either in vision or audio. Recent efforts have been exploring the research gaps for audio-video joint generation. MM-diffusion \cite{ruan2023mmdiffusionlearningmultimodaldiffusion} is a notable effort to align audio-video joint generation by using a dual U-Net with cross-attention between audio and video pairs, later being improved by \cite{sun2024mmldmmultimodallatentdiffusion}, which enhanced the training sequence by proposing a framework to perform diffusion on aligned latent representations of audio video. Other techniques such as Any-to-Any \cite{tang2023anytoanygenerationcomposablediffusion} project the latent encoding of any input modality to a shared multimodal latent space, which has the drawback of misalignment between actions and audio. Another approach to this task is the Diffusion Transformer \cite{peebles2023scalablediffusionmodelstransformers} which has shown impresive performance. Building on this work, AV-DiT \cite{wang2024avditefficientaudiovisualdiffusion} proposes using a pre-trained image transformer with a small number of trainable parameters that are adapted for image and audio generation.

The evolution of this research area leads to potential opportunities for experiments and improvements on efficient training. Some limitations on such advancements on the field include very high computational requirement and unconditional sampling, leaving a door open for improvement. Inspired by the exciting research in the field the following contributions and experiments are evaluated to expand the existing research.

\begin{itemize}
\item I release two new high-quality paired audio-video datasets: a video-game 
    gameplay dataset (13 hours of samples segmented into consistent 34-second clips) 
    and a musical concerts dataset (64 hours of samples segmented into consistent 
    34-second clips), both available upon request for non-commercial research purposes. These contribution aim at encouraging the research community to continue experimenting with joint audio-video generation.
\item{I have experimented with joint video-audio latent diffusion by employing a pretrained video encoder-decored and a pretrained audio encoder-decoder finding inconsistencies and challenges on the joint audio-video decoding step in contrast to previous techniques}
\item{I Performed a training from scratch on my new released datasets using the MM-Diffusion architecture showing consistency on video and music generation. I evaluated the effective alignment on short sounds and quick actions such is the nature of the second dataset, video games}
\item{I propose a sequential two-step text to audio-video generation procedure by leveraging two powerful video and audio generation models. By concatenating the results from the text to video generation model \cite{yang2025cogvideoxtexttovideodiffusionmodels} to the video-to-audio \cite{cheng2025mmaudiotamingmultimodaljoint} generation model and conditioning both models on long comprehensive text prompts I demonstrate good quality generation results, furthermore leveraging  negative prompting on Audio video alignment can improve even further the task. Finally I present a collection of prompt examples that show the efficacy of this method}.
\end{itemize}
\section{Related work}

\subsection{Diffusion Probabilistic Models}
Diffusion probabilistic models are a widely adopted model for image generation and sound generation. They have as a process a forward process which maps signal to noise and a reverse process that maps noise back to signal \cite{dhariwal2021diffusionmodelsbeatgans}. This family of models have been proven to perform specially well on image inpainting \cite{lugmayr2022repaintinpaintingusingdenoising}, super-resolution \cite{qiu2022learningspatiotemporalfrequencytransformercompressed}, image-to-image translation \cite{saharia2022paletteimagetoimagediffusionmodels} among others. Improving on this mechanism Denoising Diffusion implicit models \cite{song2022denoisingdiffusionimplicitmodels} are proposed as a method of sampling through a DPM in an implicit way improving sampling speed.

\subsection{video generation}
Video generation has seen an exponential improvement in recent years, with the advent of high-definition diffusion-based frameworks \cite{ho2022imagenvideohighdefinition, villegas2022phenakivariablelengthvideo}. Starting with a foundational work done by \cite{blattmann2023stablevideodiffusionscaling} who present Stable Video Diffusion (SVD), a latent video diffusion model for high-resolution text-to-video and image-to-video generation that builds upon foundational Latent Diffusion Models (LDMs) \cite{rombach2022highresolutionimagesynthesislatent} and extends prior video adaptations such as Align Your Latents. Their three-stage training pipeline comprises a text-to-image pre-training using the Stable Diffusion v2.1 image LDM \cite{blattmann2023stablevideodiffusionscaling}, large-scale video pre-training on curated Large Video Datasets (LVD)  via a systematic data curation workflow—including optical-flow-based motion filtering and OCR-based text removal and finally high-quality video fine-tuning on smaller, meticulously filtered datasets  Architecturally, SVD augments the U-Net backbone with temporal convolutional layers and temporal self-attention modules to model spatiotemporal dynamics in the latent space, while retaining cross-attention conditioning on CLIP text embeddings. The authors leverage continuous-time denoising diffusion objective with classifier-free guidance using a linear guidance schedule to balance sample fidelity and diversity.

Achieving state of the art results, \cite{yang2025cogvideoxtexttovideodiffusionmodels} propose CogVideoX, a large-scale text-to-video generation model that leverages a diffusion transformer architecture to synthesize 10-second videos at 16 frames per second with a resolution of 768×1360 pixels. To address the challenges of modeling high-dimensional video data, the authors introduce a 3D Variational Autoencoder (VAE) that compresses video inputs along both spatial and temporal dimensions. This compression facilitates efficient representation learning while preserving video fidelity. The VAE employs causal convolutions to maintain temporal coherence, ensuring that each frame generation depends only on past and present information, similar to autoregressive models in natural language processing. To enhance text-video alignment, the model incorporates an expert transformer equipped with expert adaptive LayerNorm, enabling deep fusion between textual and visual modalities.

\subsection{Sound generation}

Recent advancements in audio generation have been marked by the development of models such as AudioLDM \cite{liu2023audioldmtexttoaudiogenerationlatent}, MusicGen \cite{copet2024simplecontrollablemusicgeneration}, and AudioGen \cite{kreuk2023audiogentextuallyguidedaudio}, each introducing novel methodologies to enhance audio synthesis quality and controllability. AudioLDM \cite{liu2023audioldmtexttoaudiogenerationlatent} employs latent diffusion models trained on CLAP embeddings to generate audio from text descriptions, achieving high-quality outputs with reduced computational demands. MusicGen \cite{copet2024simplecontrollablemusicgeneration} utilizes a single-stage transformer language model operating over compressed discrete music representations, enabling efficient and controllable music generation conditioned on textual or melodic inputs. AudioGen \cite{kreuk2023audiogentextuallyguidedaudio} adopts an autoregressive transformer framework that generates audio samples conditioned on text inputs, leveraging discrete audio representations and classifier-free guidance to improve fidelity and adherence to textual prompts. Collectively, these models represent significant strides in the field of text-to-audio generation, offering diverse approaches to synthesizing audio content from textual descriptions.

These methods rely on efficient encoding in the latent space, improving on this task, there have been improvements such as Music2latent \cite{pasini2024music2latentconsistencyautoencoderslatent}, a novel end-to-end consistency autoencoder for latent audio compression that encodes complex-valued STFT spectrograms into a highly compressed continuous latent space and enables single-step high-fidelity reconstruction through a consistency model trained with a unified loss function. The architecture comprises an encoder that projects input spectrograms into a low-dimensional latent representation, a decoder that upsamples these latents back to the original resolution, and cross-connections at all hierarchical levels that condition the consistency model on intermediate encoder outputs to preserve fine-grained detail.

\begin{figure*}[t]
    \centering
    \includegraphics[height=3cm,width=\textwidth]{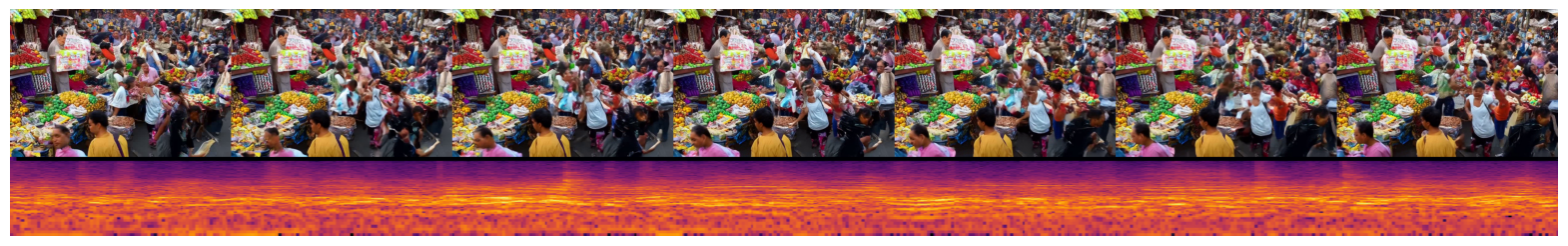}
    \caption{Two step generation with prompt: "A crowded street market in a vibrant city, filled with stalls of colorful fruits and handmade goods. Vendors shout out their prices, while...". A clear alignment between the expected noise and the video can be observed.}
    \label{fig:fullwidth_image}
\end{figure*}

\subsection{Audio-video generation}
Joint alignment on audio video generation has seen promising improvements. \cite{ruan2023mmdiffusionlearningmultimodaldiffusion} introduce MM-Diffusion, the first framework for joint audio-video generation that synthesizes semantically aligned audio–video pairs from pure Gaussian noise. The core architecture is a sequential multi-modal U-Net, comprising two coupled denoising autoencoders, one for audio spectrograms and one for video frames, that iteratively denoise noisy inputs at each diffusion timestep in lockstep. The model defines independent forward processes for audio and video with a shared linear noise schedule and employs a unified reverse model to predict noise residuals conditioned on both audio at time t and video at time t. The authors use an MM-Block, which consists on a  1D dilated convolutions for audio and decomposed 2D+1D spatial–temporal convolutions for video, interleaved with modality-specific self-attention and cross-connections to preserve fine-grained features. To enforce cross-modal coherence, the authors propose a random-shift based attention block that bridges the audio and video subnets, enabling each modality’s denoising process to attend dynamically to the other and thus reinforcing temporal and semantic fidelity across video and audio.

Building on this contribution, \cite{sun2024mmldmmultimodallatentdiffusion} present a model that extends the MM-Diffusion methodology but employs a hierarchical autoencoder to construct modality-specific low-level perceptual latent spaces that are perceptually equivalent to raw signals but significantly reduce dimensionality, alongside a shared high-level semantic feature space to bridge the information gap between modalities.

Concurrent work use alternatives to diffussion processes, such is the case of \cite{zhao2025uniformunifiedmultitaskdiffusion} which introduce UniForm, a unified multi-task diffusion transformer that jointly generates audio and video by concatenating auditory and visual tokens within a shared latent space, thereby implicitly learning cross-modal correlations and ensuring synchronous denoising across modalities. Architecturally, the model replaces dual U-Net backbones with a single transformer-based backbone featuring shared weights, which reduces parameter redundancy and enhances the exploitation of intrinsic audio-visual synergies.

Finally, there have been other approaches to deal with multimodal generation, particularly video to audio. MMAudio \cite{cheng2025mmaudiotamingmultimodaljoint} introduces a novel multimodal joint training framework designed to synthesize high-quality, temporally synchronized audio from video inputs, optionally guided by textual prompts. Departing from traditional single-modality training paradigms, MMAudio leverages extensive audio-text datasets alongside video-audio pairs to cultivate a unified semantic space, thereby enhancing the richness and diversity of audio generation. Central to its architecture is a multimodal transformer that integrates video, audio, and text modalities, augmented by a conditional synchronization module that aligns video frames with audio latents at the frame level, ensuring precise temporal coherence. The model employs a flow matching objective during training, facilitating efficient and effective learning across modalities.
\section{Methodology}

\subsection{Diffusion Models}

Diffusion models \cite{ho2020denoisingdiffusionprobabilisticmodels} are generative frameworks that learn to reverse a gradual noising process applied to data. The forward process incrementally adds Gaussian noise to the data over $T$ timesteps, transforming a clean sample $\mathbf{x}_0$ into a noisy sample $\mathbf{x}_T$. This process is defined as:

\begin{equation}
q(\mathbf{x}_t | \mathbf{x}_{t-1}) = \mathcal{N}(\mathbf{x}_t; \sqrt{1 - \beta_t} \mathbf{x}_{t-1}, \beta_t \mathbf{I}), \quad t \in [1, T],
\end{equation}

where $\beta_t$ denotes the variance schedule controlling the noise magnitude at each timestep \cite{ho2020denoisingdiffusionprobabilisticmodels}. The reverse process aims to denoise $\mathbf{x}_t$ back to $\mathbf{x}_0$ by learning a parameterized model $p_\theta(\mathbf{x}_{t-1} | \mathbf{x}_t)$, typically modeled as:

\begin{equation}
p_\theta(\mathbf{x}_{t-1} | \mathbf{x}_t) = \mathcal{N}(\mathbf{x}_{t-1}; \mu_\theta(\mathbf{x}_t, t), \Sigma_\theta(\mathbf{x}_t, t)).
\end{equation}

Where $\mu_{\theta}$ denotes the Gaussian mean value predicted by $\theta$. Training involves minimizing the variational bound on negative log-likelihood, which simplifies to the denoising score matching objective:

\begin{equation}
\mathcal{L}_{\text{simple}} = \mathbb{E}_{\mathbf{x}_0, \epsilon, t} \left[ \left\| \epsilon - \epsilon_\theta(\sqrt{\bar{\alpha}_t} \mathbf{x}_0 + \sqrt{1 - \bar{\alpha}_t} \epsilon, t) \right\|_2^2 \right],
\end{equation}

where $\epsilon \sim \mathcal{N}(0, \mathbf{I})$ and $\epsilon_\theta$ predicts the added noise.

\subsection{Multi-Modal Diffusion Model (MM-Diffusion)}

The architecture proposed in \cite{ruan2023mmdiffusionlearningmultimodaldiffusion} has shown strong performance on multi-modal audio-video joint generation. This architecture, uses two U-Net coupled denoising autoencoders. Different data types necessitate different handling on term of distributions and ensuring cross-modal coherence. Given paired data $(\mathbf{a}, \mathbf{v})$ from 1D audio set $\mathcal{A}$ and video set $\mathcal{V}$, the forward processes for each modality are defined independently:

\begin{equation}
q(\mathbf{a}_t | \mathbf{a}_{t-1}) = \mathcal{N}(\mathbf{a}_t; \sqrt{1 - \beta_t} \mathbf{a}_{t-1}, \beta_t \mathbf{I}),
\end{equation}

\begin{equation}
q(\mathbf{v}_t | \mathbf{v}_{t-1}) = \mathcal{N}(\mathbf{v}_t; \sqrt{1 - \beta_t} \mathbf{v}_{t-1}, \beta_t \mathbf{I}),
\end{equation}

with a shared noise schedule $\{\beta_t\}_{t=1}^T$ to maintain synchronization. To model the interdependence during generation, a unified reverse process is employed. Instead of separate decoders, a joint model $p_{\theta_{av}}$ conditions the denoising of each modality on both audio and video latent variables:

\begin{equation}
p_{\theta_{av}}(\mathbf{a}_{t-1} | \mathbf{a}_t, \mathbf{v}_t) = \mathcal{N}(\mathbf{a}_{t-1}; \mu_{\theta_{av}}(\mathbf{a}_t, \mathbf{v}_t, t), \Sigma_{\theta_{av}}(\mathbf{a}_t, \mathbf{v}_t, t)),
\end{equation}

\begin{equation}
p_{\theta_{av}}(\mathbf{v}_{t-1} | \mathbf{v}_t, \mathbf{a}_t) = \mathcal{N}(\mathbf{v}_{t-1}; \mu_{\theta_{av}}(\mathbf{v}_t, \mathbf{a}_t, t), \Sigma_{\theta_{av}}(\mathbf{v}_t, \mathbf{a}_t, t)).
\end{equation}

The training objective minimizes the discrepancy between the predicted and true noise for both modalities:

\begin{equation}
\mathcal{L}_{\theta_{av}} = \mathbb{E}_{\epsilon \sim \mathcal{N}(0, \mathbf{I})} \left[ \lambda(t) \left\| \epsilon - \epsilon_{\theta_{av}}(\mathbf{a}_t, \mathbf{v}_t, t) \right\|_2^2 \right],
\end{equation}

where $\lambda(t)$ is a weighting function that can emphasize certain timesteps. This joint modeling approach ensures that the generated audio and video are not only individually coherent but also semantically aligned, capturing the intricate correlations inherent in multi-modal data.

\subsection{Sequential procedure for Audio-Video Generation}

In order to perform text to audio-video conditional generation I leverage two large pre-trained models for video generation \cite{yang2025cogvideoxtexttovideodiffusionmodels} and audio to video generation \cite{cheng2025mmaudiotamingmultimodaljoint}. Starting with \textit{CogVideoX} which consists on a 3D Causal Variational Autoencoder, Latent Diffusion in Video Space and an Expert Transformer. The latent diffision on video space consist on a compressed representation $\mathbf{z}_0$ and a diffusion process which is applied in latent space. The forward (noising) process over $T$ timesteps is defined by
\begin{equation}
q(\mathbf{z}_t \mid \mathbf{z}_{t-1})
= \mathcal{N}\bigl(\mathbf{z}_t;\,\sqrt{1-\beta_t}\,\mathbf{z}_{t-1},\,\beta_t \mathbf{I}\bigr),
\quad
\bar\alpha_t = \prod_{s=1}^t (1-\beta_s).
\end{equation}
The model learns a reverse denoising network $\epsilon_\theta$:
\begin{equation}
p_\theta(\mathbf{z}_{t-1}\mid \mathbf{z}_t, \mathbf{y})
= \mathcal{N}\bigl(\mathbf{z}_{t-1};\,\mu_\theta(\mathbf{z}_t,\mathbf{y},t),\,\Sigma_\theta\bigr),
\end{equation}
with the objective
\begin{equation}
\mathcal{L}_{\mathrm{diff}}
= \mathbb{E}_{\mathbf{z}_0,\epsilon,t}\Bigl[\big\|\epsilon - \epsilon_\theta(\sqrt{\bar\alpha_t}\mathbf{z}_0 + \sqrt{1-\bar\alpha_t}\epsilon,\mathbf{y},t)\big\|_2^2\Bigr],
\end{equation}
where $\mathbf{y}$ denotes conditioning signals (text embeddings and expert controls). The second important contribution of \textit{CogVideoX} \cite{yang2025cogvideoxtexttovideodiffusionmodels}.  consist on an Expert Transformer module. \emph{Expert-specific} adapters are included  into Transformer layers via Expert-Adaptive Layer Normalization (EA-LN). Given a hidden representation $\mathbf{h}_\ell$ at layer $\ell$ and expert weights $(\gamma_e,\beta_e)$ for expert $e$, EA-LN computes:
\begin{equation}
\mathrm{EA\text{-}LN}(\mathbf{h}_\ell; e) 
= \gamma_e \odot \frac{\mathbf{h}_\ell - \mu(\mathbf{h}_\ell)}{\sigma(\mathbf{h}_\ell)} + \beta_e.
\end{equation}
Experts are activated via gating based on user-provided control tokens, enabling the same backbone to adapt dynamically to different modalities or control inputs.

The next sequential step on this framework consist on prompting a pre-trained video to sound model guided by text. MMAudio  \cite{cheng2025mmaudiotamingmultimodaljoint} is selected as a state of the art model on this task. This model uses \textit{conditional flow matching} (CFM), a continuous-time generation paradigm. The model learns a conditional velocity field $v_\theta(t, C, x)$ over the latent space $\mathbb{R}^d$, where $C$ denotes conditioning inputs (e.g., video or text), and $x$ denotes audio latent features. The sample $\mathbf{x}_1$ is obtained by integrating from Gaussian noise $\mathbf{x}_0 \sim \mathcal{N}(0, \mathbf{I})$ as:
\begin{equation}
\frac{d\mathbf{x}_t}{dt} = v_\theta(t, C, \mathbf{x}_t), \quad t \in [0,1].
\end{equation}
During training, the model minimizes the following flow-matching objective:
\begin{equation}
\mathcal{L}_{\text{CFM}} = \mathbb{E}_{t, \mathbf{x}_0, \mathbf{x}_1, C}\left[\left\|v_\theta(t, C, \mathbf{x}_t) - u(\mathbf{x}_t \mid \mathbf{x}_0, \mathbf{x}_1)\right\|_2^2\right],
\end{equation}
where $\mathbf{x}_t = t\mathbf{x}_1 + (1-t)\mathbf{x}_0$ is a linear interpolation, and $u(\mathbf{x}_t \mid \mathbf{x}_0, \mathbf{x}_1) = \mathbf{x}_1 - \mathbf{x}_0$ is the flow velocity.

\subsubsection{Sequential generation} 

Given a long-form textual prompt $\mathbf{p}$ and classifier-free guidance scale $\omega$, the video generation stage synthesizes a sequence of frames $\mathbf{v}_{1:T}$ in latent space:
\begin{equation}
\mathbf{v}_{1:T} = \texttt{CogVideoXPipeline}(\mathbf{p}; \omega).
\end{equation}
These frames are rendered and exported to a video file $\mathcal{V}_{\text{out}}$ using a fixed frame rate (e.g., 8 fps). In the second stage the \texttt{MMAudio} model is used for the task of video to audio generation. The model conditions on both visual inputs (clip-level and frame-aligned features extracted via Synchformer) and textual prompts, and produces audio latent trajectories guided by a learned velocity field. Specifically, given conditioning information $C = \{\mathbf{v}, \mathbf{p}\}$ from the video $\mathbf{v}$ and prompt $\mathbf{p}$, and a flow-matching objective $u(\cdot)$, the model integrates:
\begin{equation}
\mathbf{x}_t = \mathbf{x}_0 + \int_0^t v_\theta(\tau, C, \mathbf{x}_\tau) d\tau, \quad \text{with} \quad v_\theta \approx u(\mathbf{x}_t \mid \mathbf{x}_0, \mathbf{x}_1),
\end{equation}
where $\mathbf{x}_0 \sim \mathcal{N}(0, \mathbf{I})$ and $\mathbf{x}_1$ is the target latent. The generated audio latents are subsequently decoded into mel-spectrograms and vocoded into waveforms using a BigVGAN-based decoder, resulting in high-fidelity audio with frame-level synchronization to the video stream.
\section{Experiments}

In this section a detailed specification on the experimental setup is revised.  

\subsection{Datasets}

In order to have an effective training sequence, a high quality dataset is required for the task of joint audio video generation. Filtering by the categories "concerts" and "Call of duty" present on the Youtube 8M \cite{abuelhaija2016youtube8mlargescalevideoclassification} dataset, I created a crawling job to retrieve different videos hosted on youtube. Further processing of this information characterizes the videos by cropping each video into consistent 34 second clips and getting rid of introductory parts that could have non-related content to the video. Further inspection on the patterns of audio to filter videos that are not related to the category was performed. 

\begin{figure}[h]
    \centering
    \begin{subfigure}[t]{0.2\textwidth}
        \includegraphics[width=\linewidth]{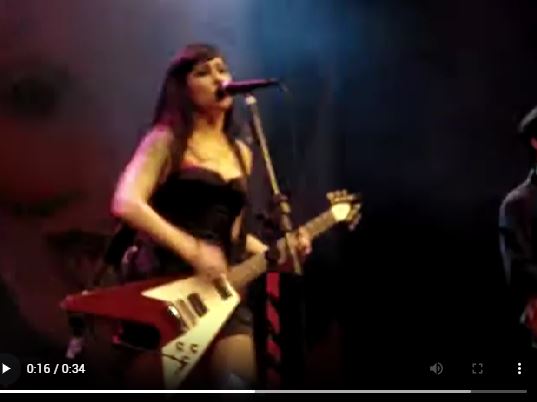}
        \caption{Concerts dataset}
        \label{fig:loss_train}
    \end{subfigure}
    \hspace{0.05\textwidth}
    \begin{subfigure}[t]{0.2\textwidth}
        \includegraphics[width=\linewidth]{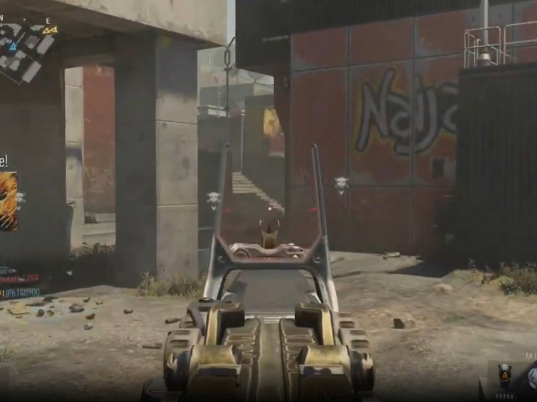}
        \caption{Gaming dataset}
        \label{fig:loss_val}
    \end{subfigure}
    \caption{Side-by-side comparison of datasets: Concerts and Gaming.}
    \label{fig:loss_comparison}
\end{figure}

As a result the concerts dataset has more than 60 hours of video audio recordings, consisting on 7200 clips, whereas the dataset of Call of Duty (i.e. gaming) resulted on 1700 clips and up to 13 hours of high quality video recordings.

\subsection{Training MM-Diffusion from scratch}

To evaluate the robustness and generalization capabilities of the MM-Diffusion framework \cite{ruan2023mmdiffusionlearningmultimodaldiffusion}, I conducted training from scratch on a newly curated audio-visual dataset. The MM-Diffusion architecture comprises two coupled U-Net-based denoising autoencoders, facilitating the joint generation of temporally aligned audio and video sequences. This design enables the model to learn cross-modal correlations effectively, ensuring coherent audio-visual outputs.

The training configuration was meticulously set to accommodate the characteristics of the custom dataset. A linear noise scheduler was employed, with the diffusion process encompassing 2000 steps. The video generation component was configured to produce sequences with dimensions $16 \times 3 \times 64 \times 64$, corresponding to 16 frames of $64 \times 64$ RGB images. Concurrently, the audio generation component was set to output waveforms with a length of 25,600 samples, aligning with the temporal span of the video sequences.

Key architectural parameters included the use of cross-attention mechanisms at resolutions 2, 4, and 8, with attention windows of sizes 1, 4, and 8, respectively. A dropout rate of 0.1 was applied to mitigate overfitting. The model utilized 128 base channels, with 64 channels dedicated to each attention head, and incorporated two residual blocks per layer. To enhance training efficiency and stability, mixed-precision training was enabled via the use of FP16 computations, and scale-shift normalization was applied throughout the network. Training was conducted using a batch size of 4 across four GPUs, with a learning rate set to 0.0001.

\begin{figure}[h]
    \centering
    \includegraphics[width=0.4\textwidth]{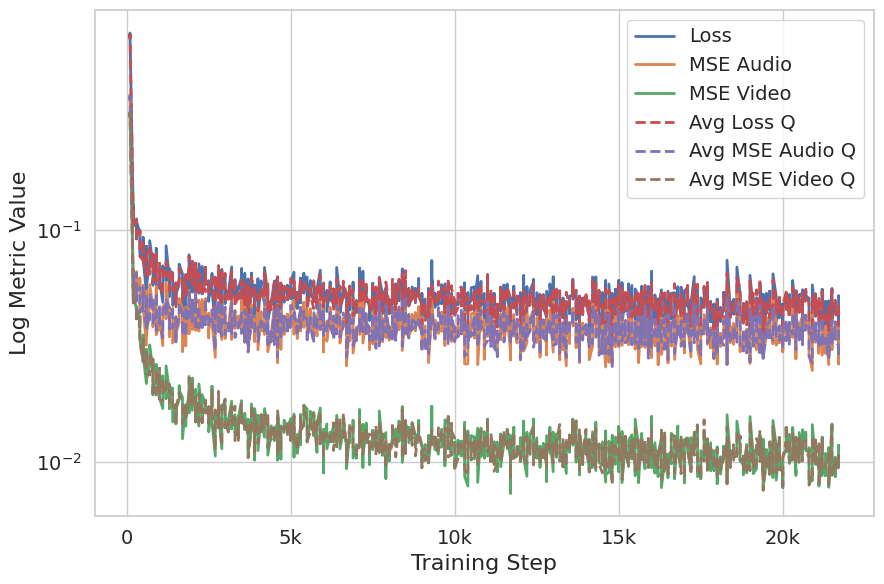}
    \caption{Loss of MM-Diffusion training with my custom concerts dataset. Listed are the coupled Joint U-net Loss, audio loss and video Loss. The model requires many resources to complete, the loss function is very noisy and requires careful hyperparameter setting.}
    \label{fig:your_label}
\end{figure}

\subsection{Latent MM-Diffusion Joint Audio video generation}
In order to improve the training time on the MM-Diffusion process, a latent diffusion model was tested. Leveraging the architecture proposed by \cite{sun2024mmldmmultimodallatentdiffusion} I experiment by employing the pretrained latent image diffusion encoder-decoder from stable diffusion \emph{sd-vae-ft-mse} \cite{Rombach_2022_CVPR} and for audio encoder-decoder I employ the pretrained audio encoder \emph{music2latent} \cite{pasini2024music2latentconsistencyautoencoderslatent}. By matching the length and dimensions of these two encoders I leverage the original MM-Diffusion training procedure to train the model. However, because of very different architectures on the decoding step, the experiments showed poor performance.

\subsection{Two-Step sequence for text to audio video generation}
For this experiment, a dataset comprising 100 prompts was generated using the LLaMA 3.2 3B language model. Each prompt, averaging 50 words in length, was designed to encapsulate detailed descriptions suitable for audio-video generation tasks. These prompts were subsequently input into the CogVideoX model , a state-of-the-art text-to-video diffusion model. CogVideoX employs a 3D Variational Autoencoder (VAE) for efficient spatiotemporal compression and an expert transformer architecture to enhance text-video alignment. The model's capabilities in generating coherent, long-duration videos with significant motion make it particularly suitable for this application.

\begin{figure}[h]
    \centering
    \includegraphics[width=\linewidth]{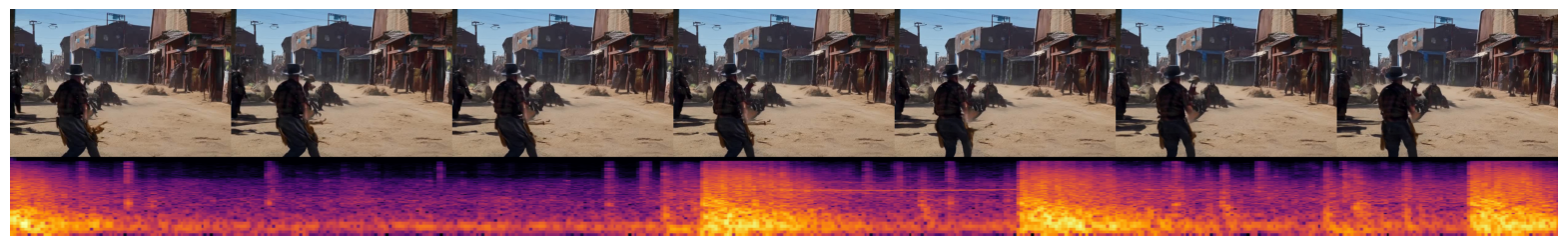}
    \caption{Two-step sequential generation using prompt "A wild west shootout at high noon in a dusty town. Spurs clink as gunslingers face off in the...".}
    \label{fig:your_label}
\end{figure}

Following video generation, each video, along with its corresponding prompt, was processed using the MM-Audio model. MM-Audio is designed for synchronized audio generation conditioned on video and/or text inputs. It utilizes a multimodal joint training framework, enabling it to learn from a diverse range of audio-visual and audio-text datasets. This approach facilitates the generation of high-quality, semantically aligned audio that maintains temporal coherence with the visual content. The integration of detailed prompts and advanced modeling techniques in both video and audio generation stages contributes to the overall effectiveness of the system.

\subsection{Evaluation metrics}
The evaluation metrics employed in this study are grounded in the Fréchet Distance (FD), which quantifies the divergence between the distributions of real and generated feature representations in a high-dimensional embedding space. Specifically, both audio and video modalities are embedded into fixed-length feature vectors via pretrained neural models—ResNet18 for spectrogram-based audio and R(2+1)D for video. For each modality, statistical properties including the empirical mean and covariance matrix of the extracted feature sets are computed. The Fréchet Distance is then calculated between these Gaussian approximations of real and generated data distributions, incorporating both mean and covariance via the closed-form solution:  
\[
\text{FD}(\mu_1, \Sigma_1, \mu_2, \Sigma_2) = \|\mu_1 - \mu_2\|^2 + \text{Tr}(\Sigma_1 + \Sigma_2 - 2(\Sigma_1 \Sigma_2)^{1/2})
\]  
This metric captures both first-order (mean) and second-order (covariance) discrepancies, offering a principled measure of the generative model’s fidelity to the real data manifold.

\begin{table}[H]
\centering
\caption{Comparison of FAD and FVD scores for unconditional and text-video generation models. For the concerts dataset. Baseline results of \cite{ruan2023mmdiffusionlearningmultimodaldiffusion}. Lower values are better.}
\label{tab:gen_results}
\begin{tabular}{lcc}
\toprule
\textbf{Method} & \textbf{FAD} ↓ & \textbf{FVD} ↓ \\
\midrule
Unconditional Generation & 9260.94 & 251.62 \\
Text-to-Video Generation & 5020.37  & 206.49 \\
\cite{ruan2023mmdiffusionlearningmultimodaldiffusion}(AIST++, dancing dataset) & 10.69 & 75.71 \\
\bottomrule
\end{tabular}
\end{table}
\section{Conclusions}

In this paper I presented the experiments performed on custom datasets using the MM-Diffusion framework for joint audio-video generation and proposed a two step sequencial procedure to generate audio-video  using CogvideoX  \cite{yang2025cogvideoxtexttovideodiffusionmodels} for video generation and MM-Audio \cite{cheng2025mmaudiotamingmultimodaljoint} for video to audio generation. The proposed implementation of latent joint audio video diffusion proved to be challenging as the decoding step becomes unstable. It is also observed that training MM-Diffusion from scratch proves to be a very rource intensive task and requires careful hyperparameter setting. Further work on on this research direction could include improving on the latent MM-Diffusion method to achieve State of the art result with less resources. 

\bibliographystyle{ieeetr}
\bibliography{references}

\onecolumn
\appendix
\section*{MM-Diffusion Trained from Scratch}
Shown below are more examples for MM-Diffusion.

\vspace{1em}

\begin{center}
\includegraphics[width=\textwidth]{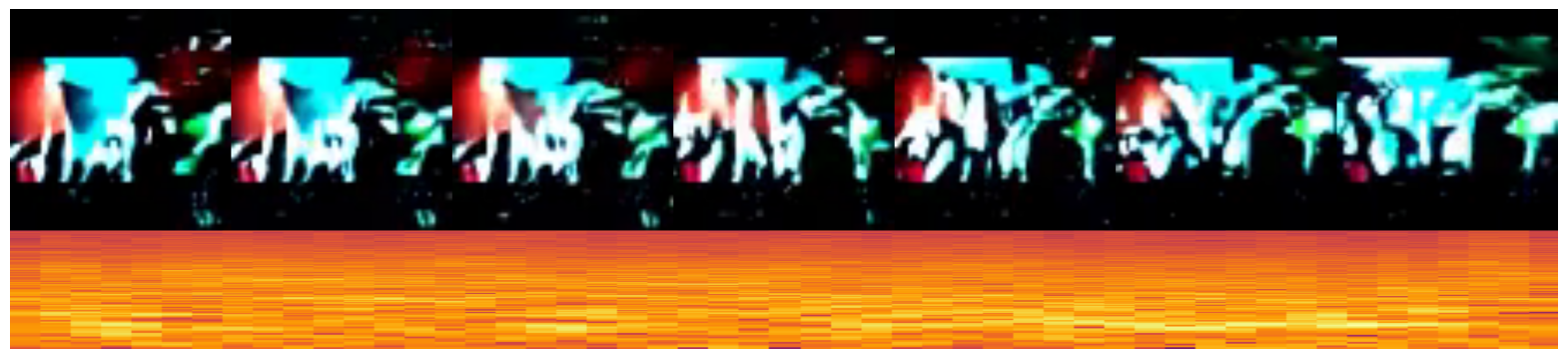}
\end{center}

\vspace{1em}

\begin{center}
\includegraphics[width=\textwidth]{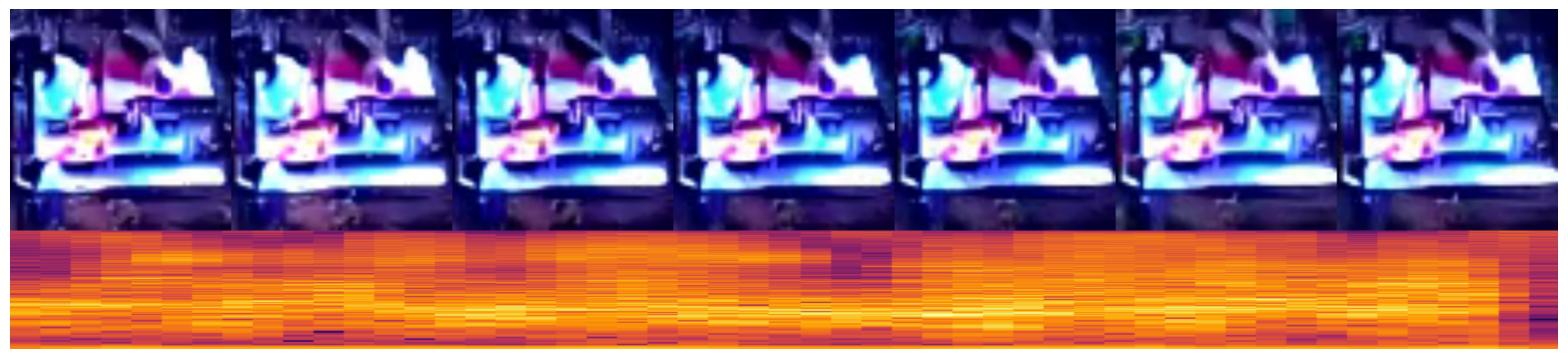}
\end{center}

\vspace{1em}

\begin{center}
\includegraphics[width=\textwidth]{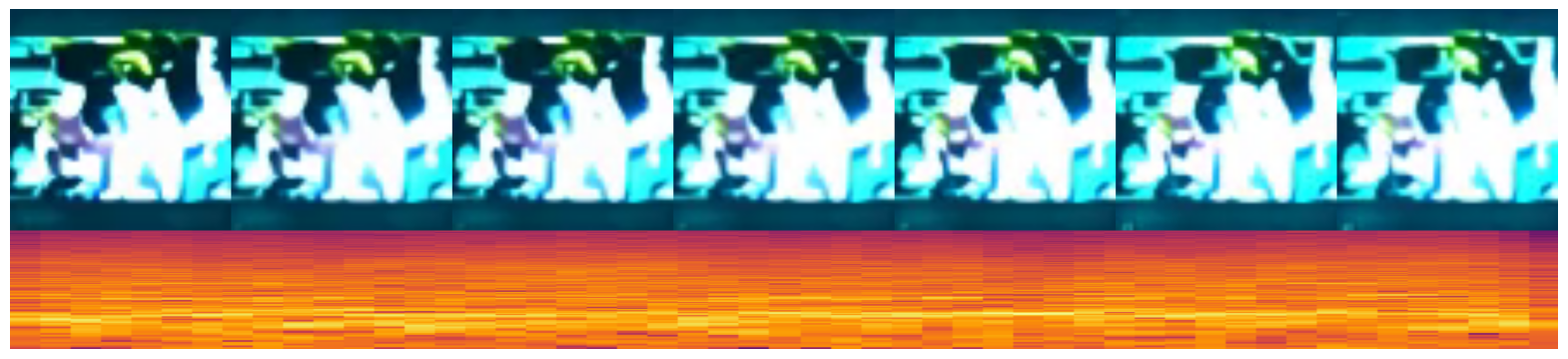}
\end{center}

\vspace{1em}
\begin{center}
\includegraphics[width=\textwidth]{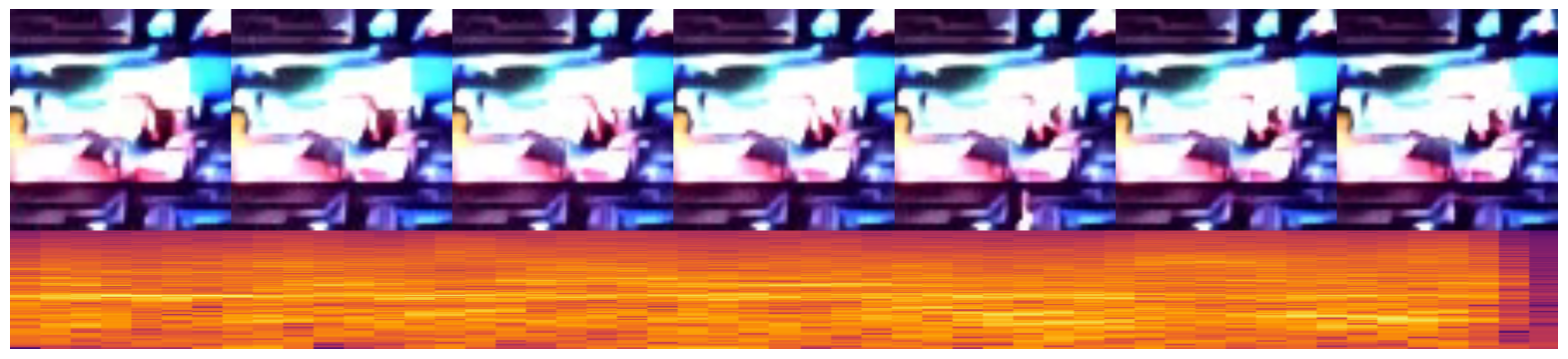}
\end{center}

\begin{center}
\textbf{Figure A1:} MM-Diffusion training from scratch on the concert dataset.
\end{center}

\newpage

\section*{Sequential audio-video generation}

\vspace{1em}
\begin{center}
\includegraphics[width=\textwidth]{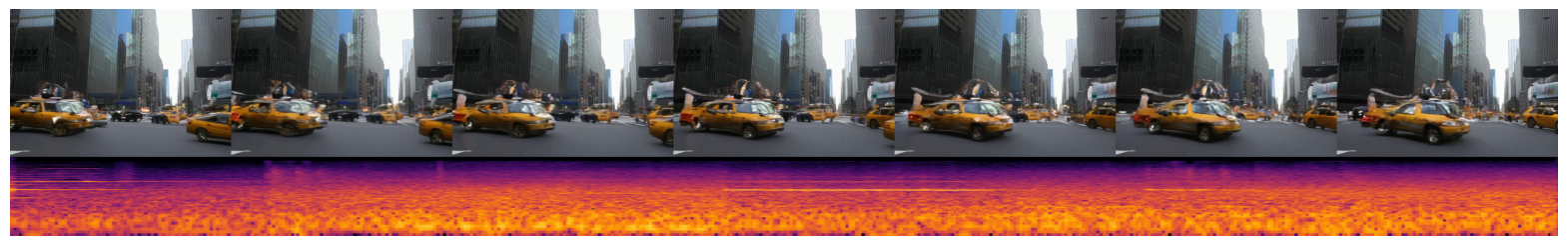}
\end{center}

\vspace{1em}
\begin{center}
\includegraphics[width=\textwidth]{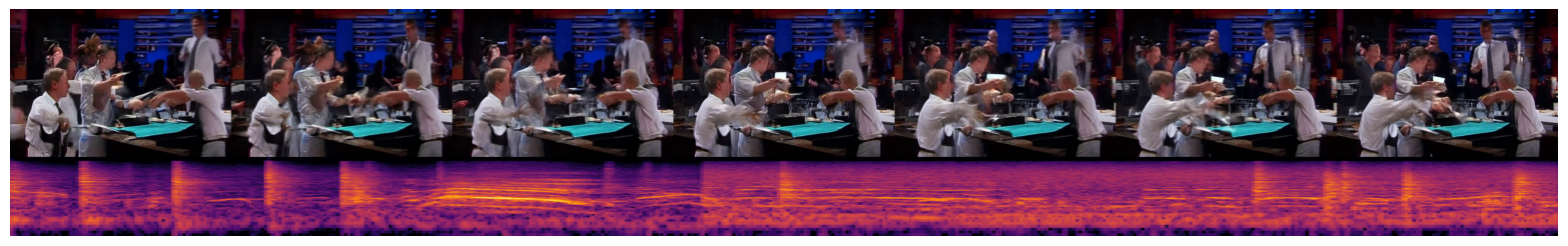}
\end{center}

\vspace{1em}
\begin{center}
\includegraphics[width=\textwidth]{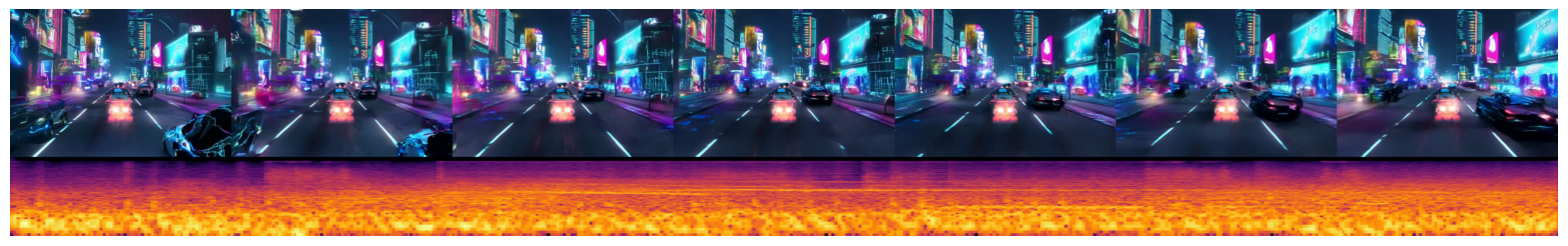}
\end{center}

\vspace{1em}
\begin{center}
\includegraphics[width=\textwidth]{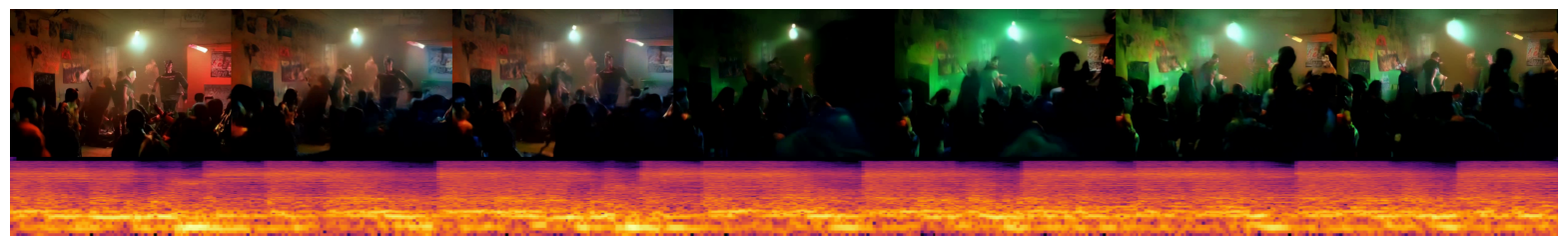}
\end{center}

\end{document}